%% file: main.tex
\documentclass[journal]{IEEEtran}
\ifCLASSINFOpdf
   \usepackage[pdftex]{graphicx}
\else
\fi
\usepackage{amsmath}
\usepackage{numprint}

\usepackage[usenames, dvipsnames]{color}

\begin{document}
%
\title{Location Order Recovery in Trails with Low Temporal Resolution}
%
%
%

\author{Binxuan Huang and
        Kathleen M. Carley,~\IEEEmembership{Fellow,~IEEE,}
\thanks{Binxuan Huang and Kathleen M. Carley are with the School of Computer Science, Carnegie Mellon University, Pennsylvania,
PA, 15213 USA e-mail: binxuanh@cs.cmu.edu; kathleen.carley@cs.cmu.edu}
\thanks{Manuscript received December 15, 2017. This work was supported in part by the Office of Naval Research (ONR) N000141512563, and the Center for Computational Analysis of Social and Organization Systems (CASOS). The views and conclusions contained in this document are those of the authors and should not be interpreted as representing the official policies, either expressed or implied, of the Office of Naval Research or the U.S. government.}}

%
%

\markboth{Journal of IEEE Transactions on Network Science and Engineering}%
{Shell \MakeLowercase{\textit{et al.}}: Bare Demo of IEEEtran.cls for IEEE Journals}
%



\maketitle
\begin{abstract}	
\input{./content/0_abstract}
\end{abstract}

\begin{IEEEkeywords}
Trail, Transition Network, Algorithm
\end{IEEEkeywords}

%
\IEEEpeerreviewmaketitle

\input{./content/1_introduction_v3}
\input{./content/2_related_v4}

\input{./content/3_method}

\input{./content/4_data}

\input{./content/5_result}

\input{./content/6_conclusion}


%

\appendices
\input{./content/7_acknowledge}

\ifCLASSOPTIONcaptionsoff
  \newpage
\fi



\bibliographystyle{IEEEtran}
\bibliography{./content/reference}

%




\begin{IEEEbiography}[{\includegraphics[width=1in,height=1.25in,clip,keepaspectratio]{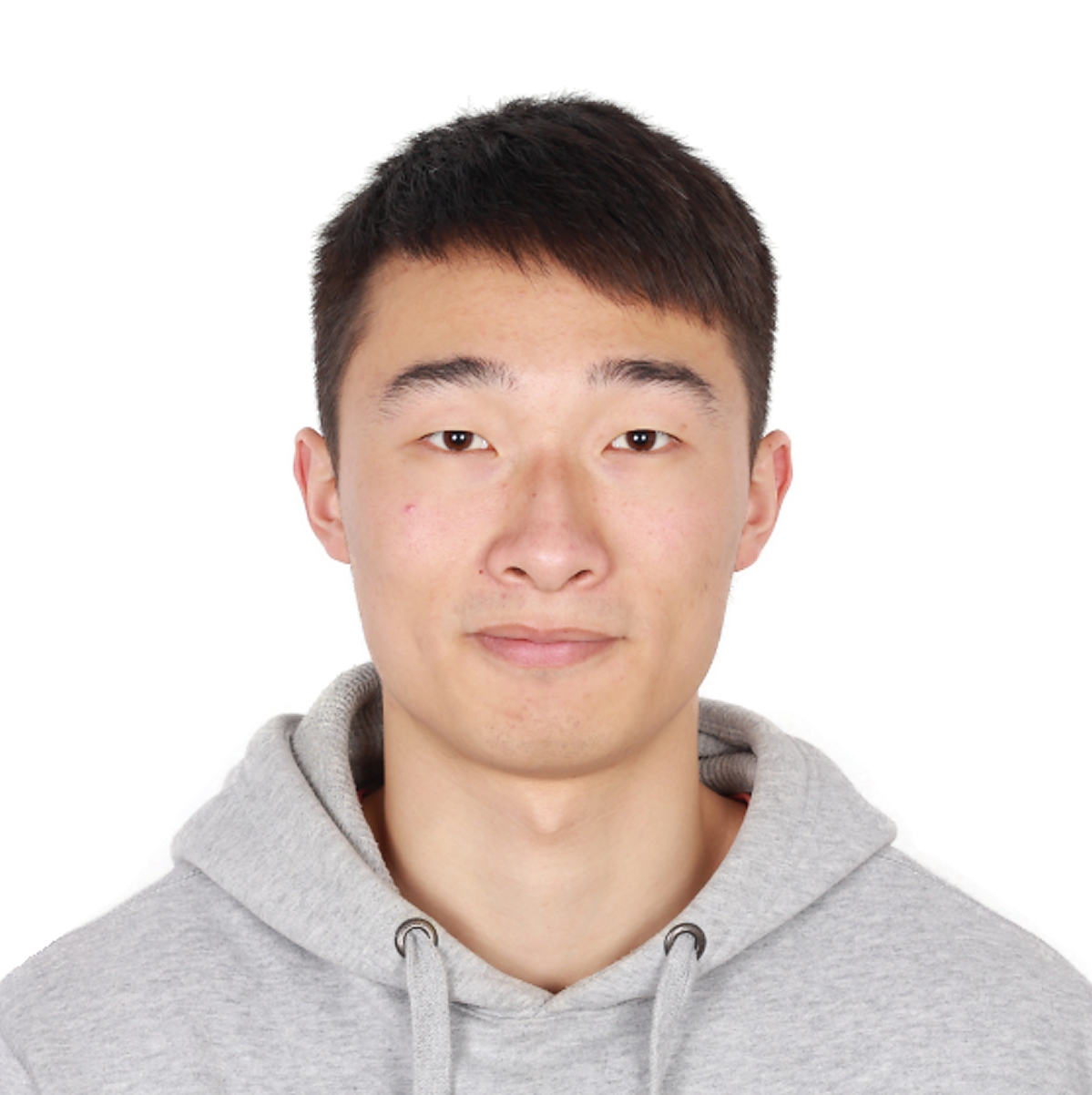}}]{Binxuan Huang}
Binxuan Huang received the B.S. degrees in physics and computer science from Zhejiang University, China in 2015. He is currently pursuing the Ph.D. degree in Societal Computing Program.

His current research interests include issues at the intersection of social network analysis and natural language processing and their application to the field of understanding people’s behaviors in online social networks.
\end{IEEEbiography}

\begin{IEEEbiography}[{\includegraphics[width=1in,height=1.25in,clip,keepaspectratio]{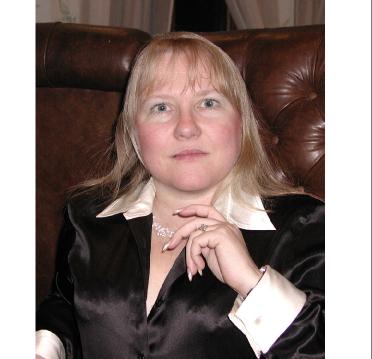}}]{Kathleen M. Carley}
Kathleen M. Carley (M’06–SM’11–F’13) received the S.B. degree in political science and the S.B. degree in economics from the Massachusetts Institute of Technology, Boston, MA, USA, and the Ph.D. degree in mathematical sociology from Harvard University, Boston, MA, USA.

She is currently a Professor of the Societal Computing Program with the Institute for Software Research Department, School of Computer Science, Carnegie Mellon University (CMU), Pittsburgh, PA, USA, where she is also the Director of the Center for Computational Analysis of Social and Organizational Systems. She holds affiliated positions and courtesy appointments with the Departments of Social and Decision Sciences, Engineering and Public Policy, and Electrical and Computer Engineering, CMU. Her current research interests include dynamic network analysis and agent-based modeling applied to issues, such as security, counterterrorism, information diffusion, and social change.

Dr. Carley is a member of the International Network for Social Network Analysis, American Statistical Association, Institute for Operations Research and the Management Sciences, and the American Association for the Advancement of Science. She received the Simmel Award in 2011 for outstanding contributions to network science, and has served on numerous National Research Council committees and IEEE Workshop committees.
\end{IEEEbiography}




\end{document}

%% file: content/0_abstract.tex
Researchers who study object movement problems related to topics like traffic flow analysis, patient monitoring, and software operation, need to know the correct order in which objects move. Here, we use the term trail to refer to a series of movements by an object. This paper introduces a new missing data problem that occurs when analyzing trails where there is inadequate temporal resolution on the events. The temporal resolution is inadequate when an object, which can only be in one place at one time, appears in the data to be in two or more locations at once.  We refer to this lack of resolution as a broken point.  Broken points prevent us from knowing the correct order of movement. We propose a three-phase framework for recovering the location order. Based on the Markov transition network, we are able to find the route with the highest probability. Our results show that this framework can efficiently find the correct location order in trails with low temporal resolution. We also demonstrate that by correcting the location order, the criticality of locations can change significantly. 


%% file: content/1_introduction_v3.tex
\section{Introduction}
People are interested in how objects move between locations. There are many research questions related to object movement, such as human mobility, traffic flow, animal migration, and so on. There are two main types of objects people are interested in: unsplittable and splittable objects. Unsplittable objects like humans\cite{gonzalez2008understanding}, animals\cite{wilcove2008going}, and vehicles\cite{lv2015traffic} can only move to one location at one time, while splittable objects like disease\cite{tatem2006global}, information\cite{gruhl2004information}, and ideas\cite{hagerstrand1968innovation} can appear at different locations at the same time. 

In this work, we focus on the location order recovery problem for unsplittable objects. Knowing the correct location order of unsplittable objects is crucial for many research questions. For example, a medical system may record all the health services one patient has visited previously. From such health records, we can learn health conditions of one patient in the past. One doctor may analyze the influence of the previous treatments on the patient. In such a situation, the health treatment order is important for a doctor's analysis. Another example is city planning. When planning city routes, people first analyze the traffic flows in the city. From previous vehicle’s movement history, researchers learned what's the popular route between important locations\cite{chen2011discovering}. Discovering the correct movement order is the first step for solving these problems.

However, research studying unsplittable object movement is built on the assumption that a location sensor can accurately record when an object visits a location with high temporal resolution. Here a location sensor is just a location recording system that records when an object visits where.
A sensor with high temporal resolution can record the current time in seconds or even microseconds, while with low temporal resolution it may only record the current time in hours or days. Because of low temporal resolution, a series of movements would appear in the same time slot, which implies that an unsplittable object visits several different locations simultaneously. The correct order of movements is missing, which can be viewed as a missing data issue. 

\begin{figure}[!h]
\centering
\includegraphics[width=3in]{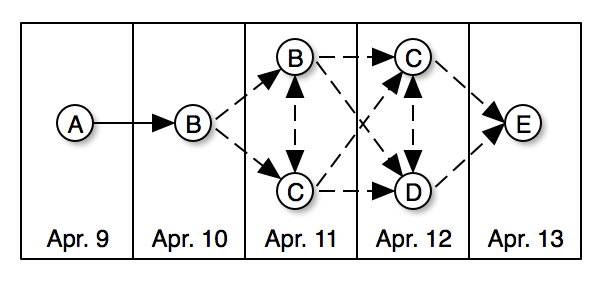}
\caption{An example of trail with low temporal resolution. One rectangle plate is a time slot in the data. Nodes are the locations an object visited in the past. A solid directed edge represents a known movement from one location to the next location. A dashed directed edge represents an undetermined potential movement.}
\label{figure1}
\end{figure}

To demonstrate this problem, we show an example of a trail in high resolution versus a trail in low resolution. Table \ref{example1} below is one trail recorded from Apr. 9 to Apr. 12. The temporal resolution in the left is at the minute level. We can clearly know that the location sequence this object visited is $A\to B\to B \to C\to C \to D\to E$. To the right is the same trail but with temporal resolution at the day level. We know that this object visited location A on April 9 and then moved to B from A on April 10. However, we can only learn that this object visited location B and C on April 11, but do not know the movement order in the time period between April 10 and April 13. Figure \ref{figure1} is a network representation of the trail in the right table. Each node in the graph is a location record. The dashed directed edges represent undetermined potential movements during April 10 and April 13. We approach this problem by finding the most probable route in the dashed network shown in Figure \ref{figure1}.

\begin{table}[!h]
\centering
\caption{Examples of trails without/with temporal resolution issue. Left is the one with higher resolution. Right is the one with different locations in one time slot.\label{example1}}{
\begin{tabular}{|l|l|l|l|l|}
\cline{1-2} \cline{4-5}
Time            & Location &  & Time      & Location \\ \cline{1-2} \cline{4-5} 
2016/4/9 10:00  & A        &  & 2016/4/9  & A        \\ \cline{1-2} \cline{4-5} 
2016/4/10 11:00 & B        &  & 2016/4/10 & B        \\ \cline{1-2} \cline{4-5} 
2016/4/11 9:30  & B        &  & 2016/4/11 & B        \\ \cline{1-2} \cline{4-5} 
2016/4/11 13:00 & C        &  & 2016/4/11 & C        \\ \cline{1-2} \cline{4-5} 
2016/4/12 10:00 & C        &  & 2016/4/12 & C        \\ \cline{1-2} \cline{4-5} 
2016/4/12 11:30 & D        &  & 2016/4/12 & D        \\ \cline{1-2} \cline{4-5} 
2016/4/13 14:00 & E        &  & 2016/4/13 & E        \\ \cline{1-2} \cline{4-5} 
\end{tabular}}
\end{table}

Even though high resolution data is often available today, there are still many cases where we cannot obtain high resolution data. A typical example is disease contamination. Patients usually don't know the exact time they got infected by some disease. As a result, it is hard to determine the transportation path for the disease without knowing the accurate infection time. Low temporal resolution is common in survey research. Respondents in a survey often cannot recall the exact time when an event happened. As a result, it is hard for a respondent to remember the correct order for a series of events that occurred even a week previously. Data storage and privacy concerns can also result in data collected at high resolution being collapsed to a lower resolution level.


In this paper, we study the problem of recovering the correct location order using the Markov transition network between locations. By considering the movement histories, our goal is to find the location order with the highest probability. Based on Markov property, we show this problem is similar to asymmetric traveling salesman problem (ATSP) in the probabilistic space. Our problem is distinct from ATSP. First, objects enter and leave the network of locations at many spots. Hence, we need to incorporate the entering and exiting information in our framework, which can help the location order recovery. Second, an object may exit the location recording system and re-enter again after a long period. In such a situation, a trail should be considered as two separate cases. Third, even though we need to find the correct order among several locations, some locations may always precedes other locations because of time constraints. In our previous example, location B always precedes location D and E. These features require a methodology that is distinct from ATSP.

To the best of our knowledge, our research is the first attempt to solve this location order recovery problem. We propose an efficient framework to recover the order of movements that consists of three distinct processing phases. We demonstrate that adding additional nodes to trails can effectively capture the starting and ending information. We prove that partitioning is a good strategy to overcome re-entry issues. Using the modified and partitioned data, we show that an exact algorithm can solve broken point problem when the number of locations in the same time slot is small.  However, this method does not scale well as the number of "simultaneous locations" increases. We then propose utilizing an ant colony system algorithm. We provide experimental results that prove the efficiency of this framework under various temporal resolutions and with broken points that vary in the number of locations per time slot.

The rest of this paper is organized as follows. In Section II, we summarize related works in studying object movements. A detailed description of our framework is given in Section III. We present the description of our dataset in Section IV. We describe our experimental setup and the corresponding results in Section V. Finally, the conclusions and the implications for future work are presented in {Section VI}.

%% file: content/2_related_v4.tex
\section{Background}

There are various ways of representing trails. One straightforward way is to treat a trail as a location sequence. Many algorithms have been proposed to find frequent subsequences in a set of trails, such as PrefixSpan\cite{pei2001prefixspan}, GSP\cite{srikant1996mining}, FreeSpan\cite{han2000freespan}. These algorithms work well in sequential pattern mining, but they are limited in that they do not take time intervals into account. 
Another way to represent trail is to transform trails into hierarchical tree structures. Hung proposed using the probabilistic suffix tree (PST) to represent a trail for each user\cite{hung2009mining}. 
 Representing a trail as a PST makes it possible to measure the similarity between two trails using editing distance. As such, similar communities based on their trails can be found.

The representation we choose is to build a transition network in which nodes represent locations and edges represent objects moving between locations. Transition networks have been used for similar but different problems\cite{chen2011discovering}. For example, in \cite{chen2011discovering}, Chen et al. tried to find the popular routes given a destination $d$. 
They used a maximum probability product algorithm to discover the popular route, which is similar with Dijkstra algorithm\cite{dijkstra1959note}. Similarly, Luo et al. designed an algorithm to find the most frequent path in a transition network by defining a $more-frequent-than\ relation$\cite{luo2013finding}. 

The research previously described is based on trails collected with high temporal resolution. In reality, there are datasets with at various levels of temporal resolution. In some of these, the resolution is not precise enough to determine the accurate order of movements. Insurance claim data for medical procedures often has this limitation.  Such low temporal resolution result in a type of missing data problem. When researchers try to assess the movements of an object of interest, obstructions such as limited storage volume, data collapse to preserve privacy, and the cost of higher resolution sensors will result in data that appears as though it is from a low resolution sensor. Missing data of some form is hard to avoid. Researchers in numerous fields have developed techniques for handling missing data\cite{stark1989high,donoho1993nonlinear}. Examples include image noise reduction\cite{pan2016image}, user attributes inference\cite{Huang2017,qian2017probabilistic}, network traffic recovery by tensor factorizations\cite{acar2011scalable} and compressive sensing signal reconstruction\cite{dai2009subspace}.  However, the specific issue of recovery of the original sequence given missing temporal resolution has not been addressed.

The missing data here is the location order resulted from low temporal resolution. Even though there are many papers trying to recover missing data in general\cite{acar2011scalable,dempster1977maximum,zhang2017rate}, most of them assume that what is missing is the data value, not the order. Hence previous methods cannot be applied directly in our problem. The closest study was done by Merrill et al. who found this low temporal resolution issue when analyzing health record data\cite{merrill2015transition}.  They used transition networks to model patients' health records. They resolved the broken point issue by determining a patient's movement opportunistically by the number of the record. This presumes that record number is a proxy for true order. Such a solution is not the optimal way to find the correct location order, and will not work across contexts. In contrast, our approach is more general.

In the next section, we will show that recovering location order under low temporal resolution is similar but not equivalent to an asymmetric traveling salesmen problem(ATSP) in the probabilistic space. The goal of ATSP is to find a Hamiltonian circuit of a graph and the path length is a minimum\cite{fischetti1992additive}. Various algorithms have been proposed to deal with ATSP. In this paper, we use the ant colony system(ACS) to deal with long location sequence with missing order. It is a distributed algorithm that uses a set of cooperating ants to find good solutions of ATSP. There are important distinctions between our location order recovery problem and ATSP. 
First, implicitly several locations are more likely to be visited at first or at last. We need to incorporate such entering and exiting information in our framework. Second, an object may exit the location recording system and re-enter again after a long period. In such a situation, a trail should be considered as two separate cases. Third, even though we need to find the correct order among several locations, some locations may always precedes other locations because of time constraints.




%% file: content/3_method.tex
\section{Method} \label{method}
\textbf{Problem definition}: Formally, a trail is a location sequence represented as $\{(l_0,t_0),$ $(l_1,t_1),...,(l_n,t_n)\}$ where $l_i$ is location and $t_i$ is timestamp, $t_0\le t_1\le ...\le t_n$.  If there exists a subsequence $\{(l_i,t_i),(l_{i+1},t_{i+1})..,(l_j,t_j)\}$ where $t_i=t_{i+1}=...=t_j$ and $l_i,l_{i+1},...,l_j$ are not the same location, then we cannot know the location order between $l_i,l_{i+1},...,l_j$. We call this trail broken. If there are more than 2 different locations appearing at one time slot, we call this time slot a broken point. It is also possible that there are multiple broken points in one trail and they can appear continuously. The goal of this paper is to recover the true order of movements in all the broken points.

In this paper, we consider an object movement as a Markov process in which the probability for an object moving to the next location is dependent on the current location. Based on this assumption, we build a Markov transition network where an edge from $A$ to $B$ represents $P(B|A)$, which is the probability of an object moving to location $B$ given current location $A$. 

Our framework can be roughly divided into three steps. The first step is adding BEGIN/END nodes as well as partitioning trails into sub-trails. 

Because here we also know the time intervals between consecutive location records, we would like to incorporate such information in our recovery framework. In a location record system, an abnormally large time interval between two consecutive location records may imply that this object once left this location system and entered it again after a long time period. For each consecutive location record pair $(l_i,t_i)$ and $(l_{i+1},t_{i+1})$, if $t_{i+1}-t_{i}$ is larger than a threshold, then we call it a gap point. We first partition these trails at each gap point. As a result, we get several separate trails from one original trail. 

In order to capture the starting and ending information in a location sequence, we add a "\_BEGIN\_" node and a "\_END\_" node at the beginning and end of each trail. As a result, the joint probability for a location sequence $l_1,l_2,...,l_n$ would change from
\begin{align}
    P(l_1)P(l_2|l_1)\times \cdots \times P(l_n|l_{n-1})
\end{align}
to 
\begin{align}
\allowdisplaybreaks
    P('\_BEGIN\_')P(l_1|'\_BEGIN\_')\times \cdots \notag \\ \times P(l_n|l_{n-1})  P('\_END\_'|l_n)
\end{align}
where $P('\_BEGIN\_')=1$.

The second step is extracting the transition probability from unbroken subsequences in the trails we got after the first step. From those unbroken subsequences, we can learn the probability edge between each location pair as
 \begin{align}P(B|A)=\frac{N(A\to B)}{N(A)}\end{align} where $N(A)$ is the total number of location $A$ in the unbroken subsequence and $N(A\to B)$ is the number of objects moving from $A$ to $B$. Based on the Markov transition probability, we can build a transition network among locations in the broken points. 

\begin{figure}[!h]
\centering
\includegraphics[width=0.4\textwidth]{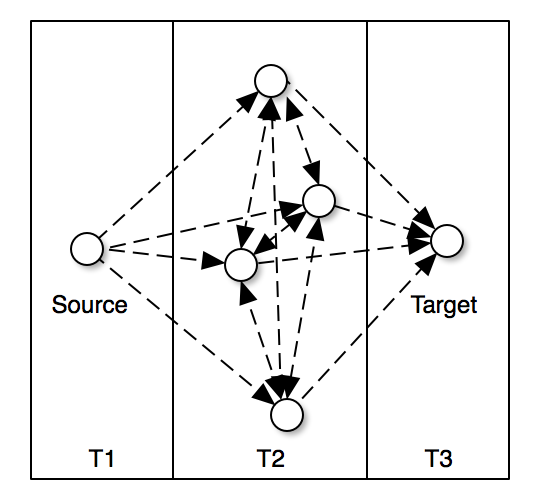}
\includegraphics[width=0.5\textwidth]{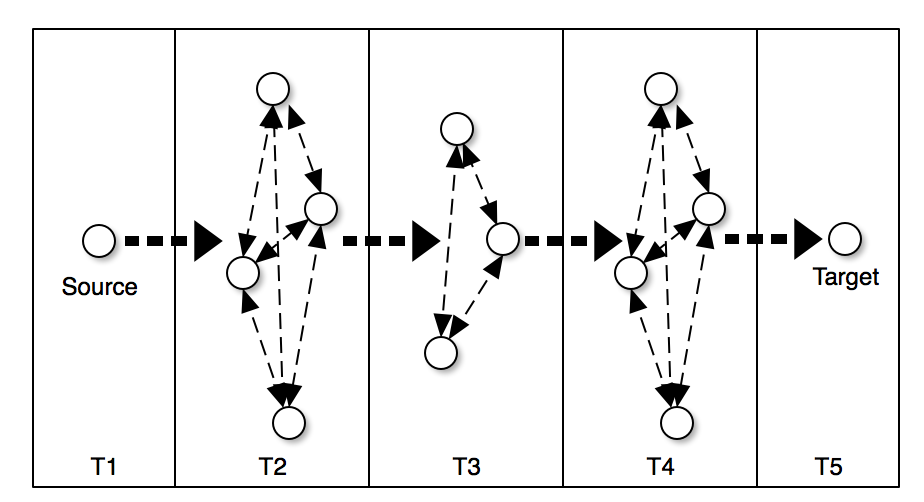}
\caption{The first one is a trail with only one broken points. Second is a trail with three continuous broken points. One object first needs to pass all the locations in layer T2, then T3 and T4. }
\label{example3}
\end{figure}

After the second step, the problem is transformed into finding the location order in a broken point $\{(l_{i},t_{i}),..,(l_j,t_j)\}$ with the highest probability product $P(l_{i-1} \to ...,\to l_{j+1}) = p(l_{i-1})p(l_{i}|l_{i-1})...p(l_j|l_{j-1})P(l_{j+1}|l_j)$, where $l_{i-1}$ and $l_{j+1}$ are locations preceding and following the broken point respectively. Basically, this is similar to an asymmetric traveling salesman problem(ATSP) which is an NP-hard problem\cite{cirasella2001asymmetric}. Because \begin{align} logP(l_{i-1} \to ...,\to l_{j+1}) = logp(l_{i-1})+\sum_{k=i-1}^{j}logp(l_{k+1}|l_k),\end{align} finding a route which passes all the locations in a single broken point with the highest probability is equivalent to an asymmetric traveling salesman problem. In our problem setting, the distance from location $A$ to $B$ is $-logP(B|A)$. When there are consecutive broken points, the problem becomes more complex. In Figure \ref{example3}, there are two examples of broken trails. One has only one broken point and the other has three continuous broken points. $Ti$ is the timestamp for the location records in the corresponding layer. In the first example, an object only needs to pass all the locations in layer $T2$, which is similar to ATSP. However, in the second example, an object first needs to pass all the locations in layer $T2$ then $T3$ and $T4$, which differs from ATSP. For simplicity, we did not draw all the possible edges between layers. The location order in $T4$ is dependent on movement history in $T2$ and $T3$. In this case, some distances between certain location pairs are infinity, eg. an object cannot move directly from layer $T2$ to $T4$ without passing $T3$ and it also cannot move in the wrong directions like from $T3$ to $T2$. 

We compared three different algorithms to find such a route with the highest transition probability. The first one is a brute force algorithm, which simply enumerates all the possible routes. Such exact algorithm can achieve the most accurate estimation of the joint transition probability. However, the time complexity is $O((N!)^T)$ where $N$ is the maximum number of locations in a broken point and $T$ is the number of continuous broken points. 

The second algorithm we use is called ant colony system(ACS)\cite{dorigo1997ant}, which is developed to deal with traditional asymmetric traveling salesman problem. As we discussed above, the distance from location A to location B is $-logP(B|A)$. Except for those determined or dashed potential edges, all the other edge distances are infinity. We use ACS algorithm to find the minimum distance in broken points. The time complexity of ACS is $O(mN^2L)$ where $m$ is the number of ants chosen manually and $L$ is the number of iterations. We used 10 ants and 300 iterations in our algorithm. Following \cite{dorigo1997ant}, we set the parameter in ACS as $\tau_0=0.5, \beta = 2, q_0 = 0.9, \alpha=0.1, \rho = 0.1$. Please refer to \cite{dorigo1997ant} for the meaning of these parameters. One thing to note is that even though we already know the true starting location and ending location in our problem, we still need to randomize the starting locations for these ants because in our experiments limiting the starting location to the source node will decrease the exploration ability of this algorithm and thus make the performance worse.

The last algorithm is a greedy algorithm, which can be viewed as a baseline for comparison. It simply starts from the source location and finds the next location with highest transition probability until arriving at the target location. The time complexity of this algorithm is $O(TN^2)$.

We also used a random strategy as a simple baseline. It randomly guess the order of locations that appears in each broken points.

\begin{figure}[!h]
\label{workflow}
\centering
\includegraphics[width=0.3\textwidth]{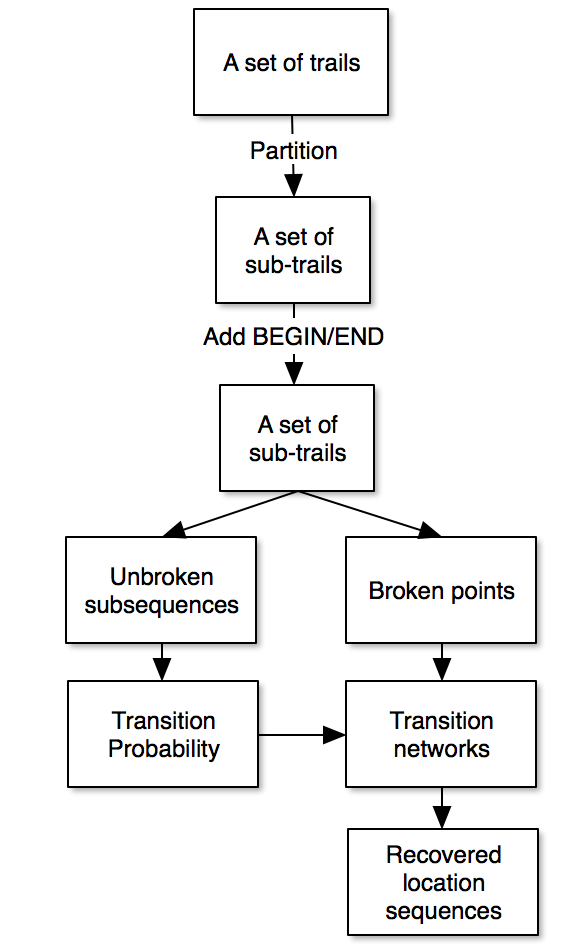}
\caption{The overall workflow of our framework.}
\end{figure}

The overall workflow of our framework is shown in Figure \ref{workflow}. To sum up, the framework consists of the following three phases:
\begin{enumerate}
\item Preparation phase: each trail is partitioned at gap points. We treat each partition as an independent trail and add BEGIN/END nodes for each partition.
\item Probability extraction phase: We extract transition probabilities from the unbroken subsequences in trail partitions we got in phase 1. Then we build Markov transition networks among locations in broken points.
\item Recover phase: find the location order with maximum transition probability.
\end{enumerate}

%% file: content/4_data.tex
\section{Dataset}
There are two datasets we used in this paper. One is data supplied by Columbia University Medical Center. It contains de-identified administrative data for patients who visited in-patient and/or out-patient service sites between July 1 2011 and June 30 2012. There are 94885 records in this data. Each record is a single visit by one patient on one specific day. The basic statistics of this dataset are shown in the Table \ref{stat_data}. The temporal resolution for the health data is at the day level.  Thus we cannot know the true location order when one patient visits several different health services on the same day. Among 5055 patients' health trails, overall, there are 4241 patients with broken health trails. Because we do not know the true location order for these broken health trails, we only used the unbroken ones to test our framework.

The other dataset is retrieved from program function call records. It is retrieved from a Java program called FindBugs\footnote{http://findbugs.sourceforge.net/}. There are 61 classes involved in this part of the program records. Each class in the program can be viewed as a location visited by the computer processor. In the original dataset, each row is a function call record in the following format:\\
\textit{call;source class; target class; method called; timestamp in nanoseconds} \\
\textit{call-end; source class; target class; method called; timestamp in nanoseconds}

This record represents the movement of the computer processor from source class to target class or return from target class. The temporal resolution is at the nanosecond level for this data, so the function trail for this program is unbroken.
\begin{table}[!h]
\centering
\caption{Basic statistics of our datasets.\label{stat_data}}{
\begin{tabular}{|l|l|l|l|l|l|}
\hline
 & \# of records & \# of agents & \# of locations &\begin{tabular}[c]{@{}l@{}}\# of\\ unbroken trails\end{tabular}  \\ \hline
 Health &94885         & 5055         & 115             & 814                   \\ \hline
Program & 69461         & 1            & 61              & 1                     \\ \hline
\end{tabular}}
\end{table}

\begin{figure}[h!]
\includegraphics[width=0.5\textwidth]{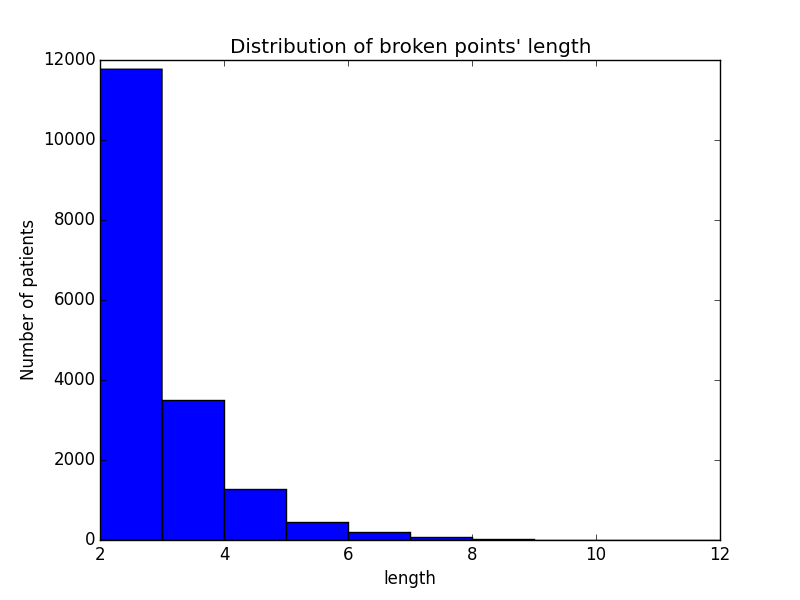}
\includegraphics[width=0.5\textwidth]{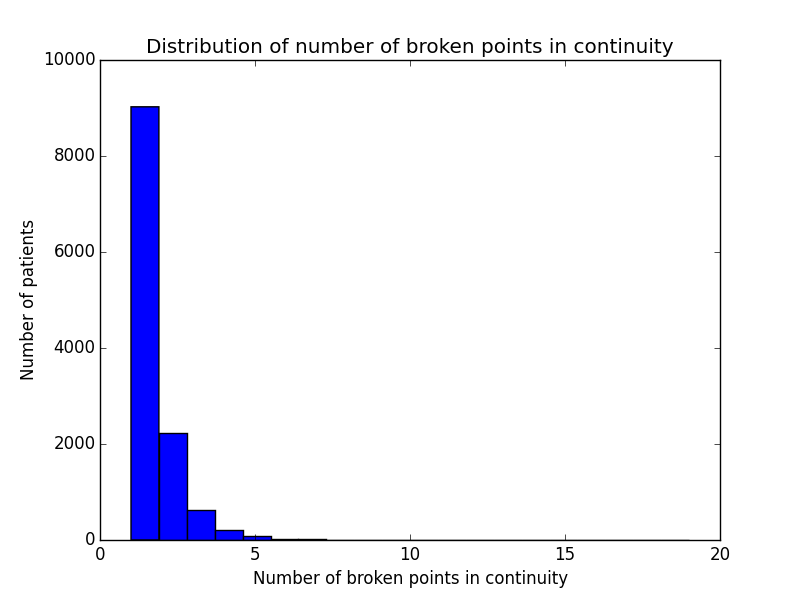}
\includegraphics[width=0.5\textwidth]{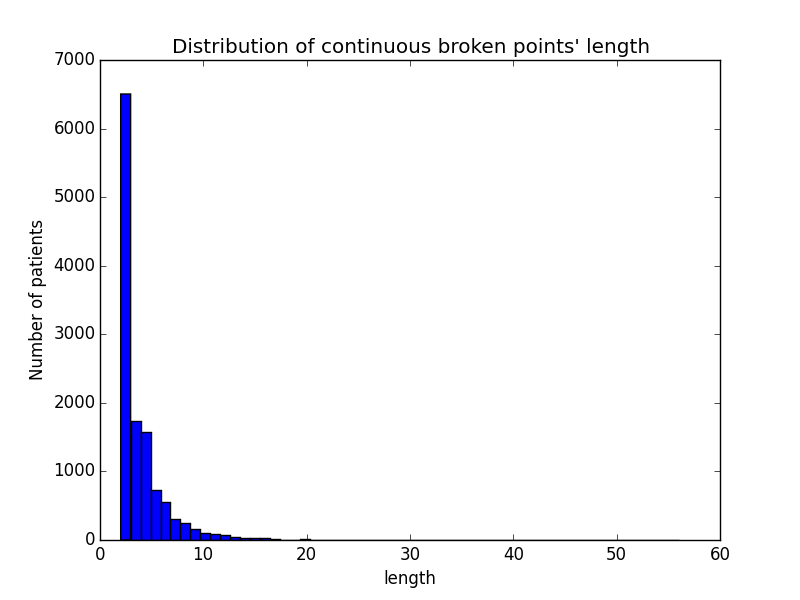}
\caption{Distributions of three factors that can affect our algorithms in the health data. They are broken point's length, number of broken points in continuity, and continuous broken points' total length.}
\label{dist}
\end{figure}

\begin{figure}[!h]
\centering
\includegraphics[width=0.5\textwidth]{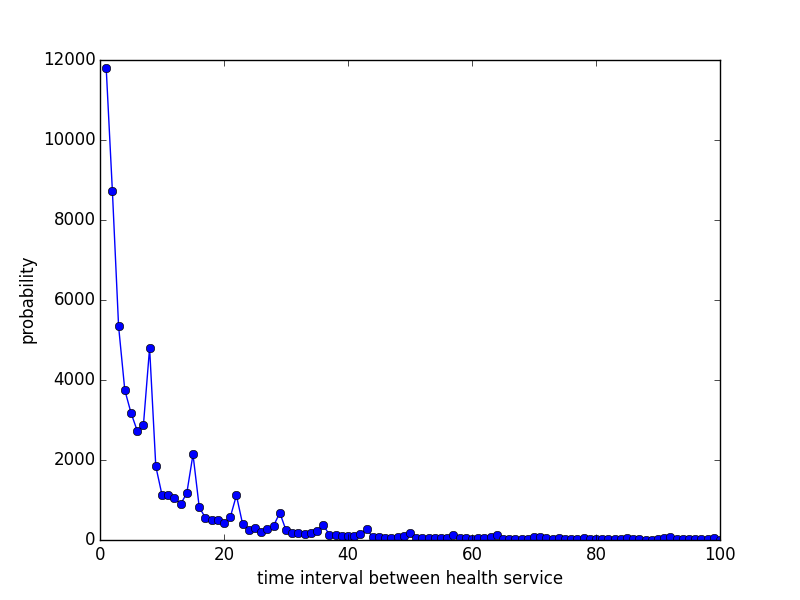}
\caption{Time interval distribution of health data, i.e. (0,11750) represents there are 11750 transitions happened within one day}
\label{interval}
\end{figure}
There are three factors that will affect the efficiency of our algorithms: the number of locations in a broken point, the number of broken points in continuity, and the total number of location records in continuous broken points. We examined these three factors in our health record data. As shown in Figure \ref{dist}, most of the broken points only contain two locations and few of them appear in continuity. The average length of broken points is 2.506 and the standard deviation is 0.921. The average length of continuous broken points is 3.534 and the standard deviation is 2.802. The average number of broken points in continuity is 1.410 and the standard deviation is 0.92. Therefore, an exact algorithm is feasible for most of the broken points whose lengths are less than 6. However, there are still few continuous broken points with length greater than 10.\footnote{Consider a single broken point with length 10, the time complexity is $10!=3628800$ which is 1814400 times higher than length 2. Given a big dataset, using an exact algorithm on these long instances would consume a lot of machine time.} When the length of continuous broken points gets larger, an approximation algorithm like ACS is a better option regarding the running time.

As shown in Figure \ref{interval}, the time interval distribution in health record data roughly follows a power law distribution. However, there are also weekly periodic patterns in the distribution. Patients tend to pay the next visit at the same weekday. Only in very rare cases is the time interval between two consecutive health services larger than one month. That is why we decided to partition health trails if the time interval between two consecutive locations was larger than 28 days.

%% file: content/5_result.tex
\section{Experiments}
\subsection{Experiments setup}
\begin{table}[!h]
\centering
\caption{Example of changing temporal resolution. On the left is an original unbroken trail. On the right is the trail with temporal resolution of two days.\label{test_example}}{
\begin{tabular}{|l|l|l|l|l|}
\cline{1-2} \cline{4-5}
\multicolumn{2}{|l|}{Original unbroken health trail} &  & \multicolumn{2}{l|}{Broken health trail after change} \\ \cline{1-2} \cline{4-5} 
7/11/2011              & Emergency                   &  & 7/11/2011               & Emergency                   \\ \cline{1-2} \cline{4-5} 
7/12/2011              & Adult Medicine              &  & 7/11/2011               & Adult Medicine              \\ \cline{1-2} \cline{4-5} 
7/15/2011              & Adult Medicine              &  & 7/15/2011               & Adult Medicine              \\ \cline{1-2} \cline{4-5} 
7/16/2011              & Geriatrics                  &  & 7/15/2011               & Geriatrics                  \\ \cline{1-2} \cline{4-5} 
7/17/2011              & Adult Medicine                  &  & 7/17/2011               & Adult Medicine                  \\ \cline{1-2} \cline{4-5} 
\end{tabular}}
\end{table}

\begin{center}
\begin{table*}[!h]
\caption{Statistics about the broken health trails we got after we changing temporal resolution of 814 unbroken health trails.\label{tb:res}}{
{\small
\hfill{}
\begin{tabular}{|l|l|l|l|l|l|l|l|l|}
\hline
Temporal resolution(day)                                                           & 2 & 3 & 4 & 5 & 6 & 7 & 8 & 9 \\ \hline
\# of broken trails                                                            & 164 & 213 &    250    & 263    &268    &278    &294    &297  \\ \hline
\# of broken points                                                            &  178    &251&    296&    319&    340&    353&    374&    389  \\ \hline
Avg. length of broken points                                                   &  2.073    &2.223&    2.284&    2.426&    2.482&    2.586&    2.591&    2.627   \\ \hline
Avg. length of cont. broken points & 2.121&    2.364&    2.449&    2.651&    2.749&    2.917&    2.910&    3.042   \\ \hline
Max. length of broken points                                                   &  4&    5&    6&    5&    6&    6&    6&    7  \\ \hline
\end{tabular}}}
\hfill{}
\end{table*}
\end{center}
In order to validate the efficiency of our location order recovery algorithms, we used the unbroken trails in our two datasets as the ground truth. We used two different strategies to create trails with artificial broken points. The first one is manually changing the temporal resolution for the unbroken trails. This can be easily achieved by $timestamp' = \lfloor\frac{timestamp}{resolution}\rfloor\times resolution$. We applied this method to health dataset. An example is shown in Table \ref{test_example}. In this example, we broke an health trail by changing its temporal resolution from one day to two days.

The original temporal resolution of the health data is one day. In our experiment, we changed the temporal resolution for all unbroken trails and used them as our test data. We broke the health trails by creating eight different low temporal resolution conditions ranging from 2 days to 9 days. Then we extracted the transition probabilities among locations from the resolution changed health trails. Table \ref{tb:res} shows the statistics of these artificial broken points under different temporal resolution.

The second test data creation strategy is called order mutation. We first determined the number of locations in a broken point as a fixed number $v$. Then we randomly selected a location visiting record $(l_i, t_i)$. After that we modified all the timestamps of the following location records between $(l_i, t_i)$ and $(l_{i+v-1}, t_{i+v-1})$ as $t_i$ and mutated the order of these records. In this way, we can create broken points with the desired sizes. To ensure we have a consistent quality of transition probability, we only randomly mutated 20\% of location records in the program function call data. In Table \ref{test_example2} is an example of trail with a broken point created by order mutation. We selected the length of broken points ranging from 2 to 15 and mutated 20\% location records in the program data. Table \ref{pro_stat} shows the number of broken points for each size.

\begin{table}[!h]
\centering
\caption{Example of order mutation for function call data. On the left is an original unbroken trail. On the right is the trail with one broken point of size three.\label{test_example2}}{
\begin{tabular}{|l|l|l|l|l|}
\cline{1-2} \cline{4-5}
\multicolumn{2}{|l|}{Original unbroken program trail} &  & \multicolumn{2}{l|}{Broken program trail after change} \\ \cline{1-2} \cline{4-5} 
Timestamp              & Java class                   &  & Timestamp               & Java class                  \\ \cline{1-2} \cline{4-5} 

553987065457672              & Class 1                   &  & 553987065457672               & Class 1                   \\ \cline{1-2} \cline{4-5} 
553987065574331              & Class 2              &  &553987065574331 & Class 3              \\ \cline{1-2} \cline{4-5} 
553987065768508             & Class 3              &  &553987065574331 & Class 4              \\ \cline{1-2} \cline{4-5} 
553987065819048              & Class 4                  &  & 553987065574331              & Class 2                  \\ \cline{1-2} \cline{4-5} 
553987100679470              & Class 1                 &  & 553987100679470              & Class 1                \\ \cline{1-2} \cline{4-5} 
553987100679860              & Class 2                 &  & 553987100679860              & Class 2                \\ \cline{1-2} \cline{4-5} 

\end{tabular}}
\end{table}

\begin{table}[!h]
\centering
\caption{Number of broken points we got after order mutation for the program data.\label{pro_stat}}{
\begin{tabular}{|l|l|l|l|l|l|l|l|l|}
\hline
broken point size       & 2    & 5    & 7    & 10   & 12   & 15  \\ \hline
number of broken points & 6810 & 2772 & 1980 & 1387 & 1155 & 925 \\ \hline
\end{tabular}}
\end{table}

We used two metrics to measure the performance of the location order recovery framework. The first one is the overall recovery accuracy which is the percentage of broken points that are recovered with the correct location order. The second one is the average hamming distance between recovered location sequences in broken points and the correct ones. It measures the minimum number of substitutions required to change one location sequence into another. We applied it to the order mutation experiment to examine the effectiveness of our framework over broken points with the same size.

\subsection{Experiment results}

\begin{figure}[!h]
\centering
\includegraphics[width=0.5\textwidth]{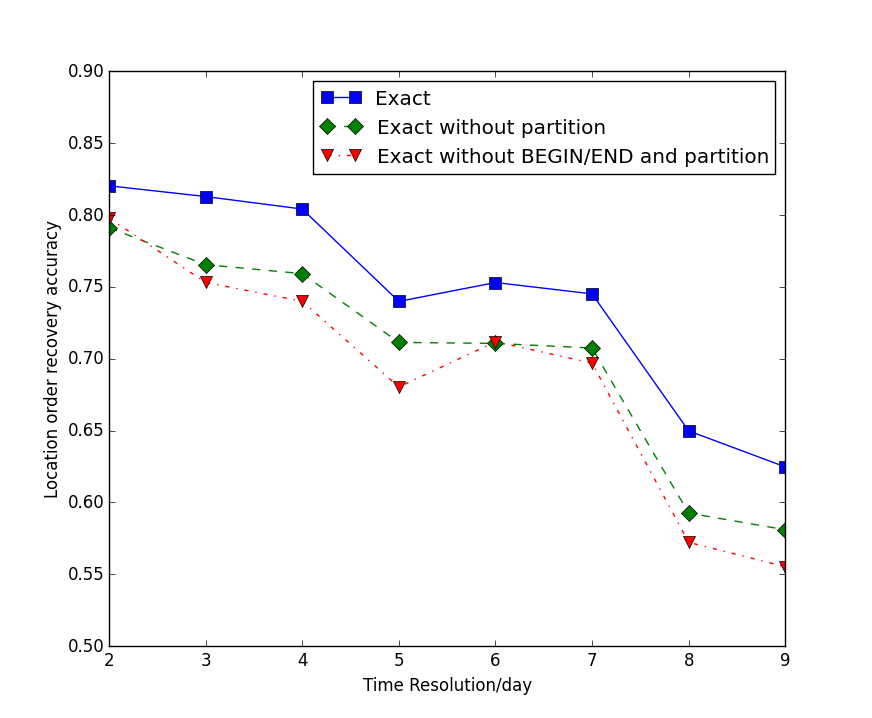}
\caption{Location order recovery accuracy for an exact algorithm with three different options: without BEGIN/END and partitioning, adding BEGIN/END, adding BEGIN/END and partitioning. }
\label{result2}
\end{figure}

We first examined the effectiveness of adding BEGIN/END locations as well as trail partitioning on the health record data. The accuracies are evaluated at each broken points, i.e. 80\% accuracy means 80 recovered broken points are correct out of 100 total broken points. In Figure \ref{result2}, we can see that when we add BEGIN/END locations to the trails, we get better recovery accuracy compared to when we do not add. This implies that adding BEGIN/END locations provides additional information for the exact algorithm when temporal resolution keeps decreasing. We also did similar experiment to evaluate the effect of including time interval factor. As shown in Figure \ref{result2}, trail partitioning would dramatically improve the location order recovery accuracy consistently. These two cases prove that simply taking the location order recovery problem as an ATSP is not an optimal solution. Adding nodes and trail partitioning would dramatically increase the performance compared with taking it as an ATSP and using an exact algorithm without these two methods. 

\begin{figure}[!h]
\centering
\includegraphics[width=0.5\textwidth]{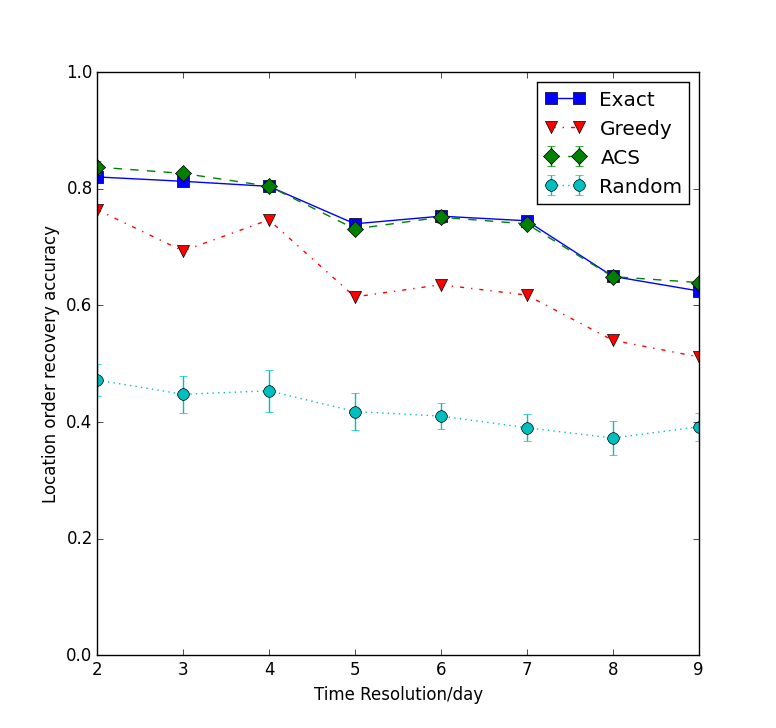}
\caption{Location order recovery accuracies for three algorithms and random guess on health data.}
\label{result1}
\end{figure}

We further compared the location order recovery performance of different algorithms. In Figure \ref{result1} are the location order recovery accuracies we got on the health trail data with BEGIN/END nodes and partitioning. 
Of the three algorithms, the exact and ACS algorithms work better and their performance is almost the same. The greedy search algorithm's accuracy is similar to the exact algorithm when the temporal resolution is at a high level. Interestingly, when we compared the accuracy for readmission patients with normal patients, we found that the recovery accuracy for readmission patients' health trails at 2-day resolution is 77.5\% while the accuracy for normal patients without readmission is 86.8\%. This suggests that normal patients have more predictable movement, which is consistent with them being either healthy and/or having accurately diagnosed and well managed conditions. 

\begin{center}
\begin{figure*}[!h]
\centering
\includegraphics[width=0.8\textwidth]{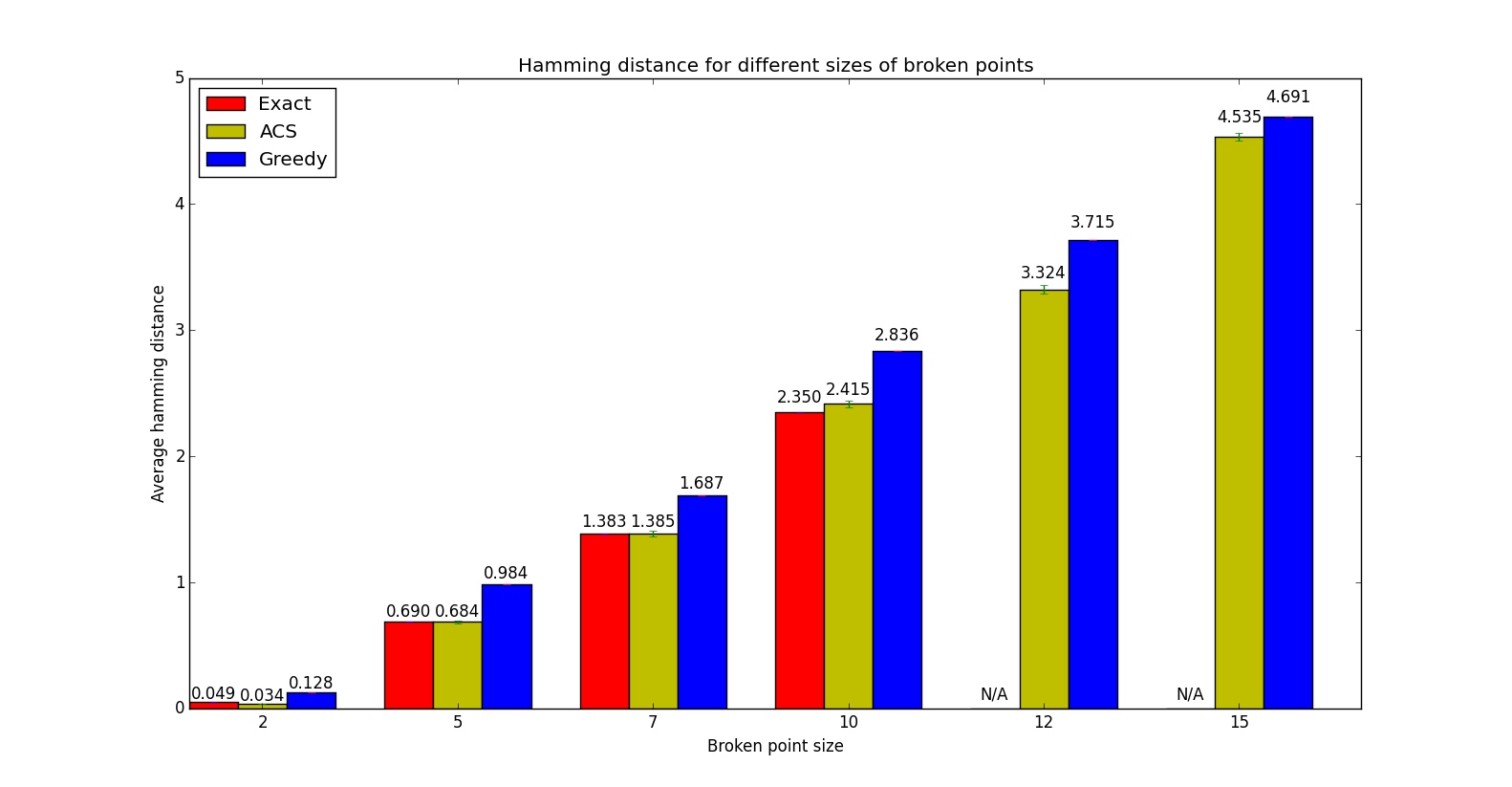}
 \caption{The average hamming distance between recovered location sequences and the true sequences for program trail.}
\label{result3}
\end{figure*}
\end{center}

To examine the effectiveness of our framework on broken points of different fixed sizes, we calculated the recovery accuracy and the average hamming distance for broken points of different sizes for the simulated trails. We only used the simulated trails in this experiment because it requires long trails, and the lengths of health trails in the test data are too short. As Table \ref{program_accuracy} shows, the accuracies of an exact algorithm and the ACS algorithm are almost the same. Sometimes ACS is better when the size of broken points is less than 12. The exact algorithm does not scale, and the time to completion prevents us from including its results for cases where the size of the  broken points is large.  We find that for larger size broken points, the accuracies of the ACS and greedy algorithms get closer. However, on examination of the average Hamming distance we found, as is shown in Figure \ref{result3},  that there is a big performance difference between the ACS and greedy algorithms.

\begin{table*}[!h]
\centering
\caption{The location order recovery accuracies we got for the program trail with order mutation. In the brackets are the standard deviation for 10 times experiments.\label{program_accuracy}}{
\npdecimalsign{.}
\nprounddigits{3}
\begin{tabular}{|c|n{1}{3}|n{1}{3}|n{1}{3}|n{1}{3}|n{1}{3}|n{1}{3}|}
\hline
broken point size & 2                                         & 5                                         & 7                                        & 10                                        & 12                                         & 15                                        \\ \hline
Exact             & 0.9754772393538913                        & 0.7301587301587301                        & 0.5727272727272728                       & 0.43258832011535686                       & N/A                                        & N/A                                       \\ \hline
ACS               & 0.9829955947136565 (0.001) & 0.7381673881673881(0.004) & 0.5928787878787879(0.008) & 0.47793799567411677(0.005) & 0.39705627705627705(0.002) & 0.3445405405405405(0.004) \\ \hline
Greedy            & 0.936123348017621                         & 0.6262626262626263                        & 0.5247474747474747                       & 0.4239365537130498                        & 0.38961038961038963                        & 0.34486486486486484                       \\ \hline
\end{tabular}\npnoround
}
\end{table*}

\subsection{Case study}
In this section, a case study using health record data is described. Knowing the sequence with which services are used in a health system is critical to improving health services, reducing costs, and improving health outcomes. One of the key question is what services are critical wayports.  In this case study we examine whether improved estimation of sequence, by resolving broken points, alters our understanding of what are the critical wayports.  We first build a directed location network, where the edge weight represents the number of patients moving from one location to another. We use inverted betweenness to measure the importance of a location. Nodes high in inverted betweenness are wayports.  Inverted betweenness is similar with standard betweenness centrality except the edge weights are inverted prior to calculation. The higher of the inverted betweenness the more important that location as a wayport.

\begin{table*}[!h]
\centering
\caption{The location ranks under different settings. The first rank is calculated from these unbroken trails before temporal resolution change. The second rank is calculated from recovered trails after resolution change while the last is calculated without recovery.}
\label{compare}{
\npdecimalsign{.}
\nprounddigits{3}
\begin{tabular}{|l|l|l|l|}
\hline
Rank & Truth                  & With recovery          & Without recovery       \\ \hline
1    & Echocardiography       & Echocardiography       & General Cardiology     \\ \hline
2    & Internal Medicine      & General Cardiology     & Electrophysiology      \\ \hline
3    & General Cardiology     & Internal Medicine      & Echocardiography       \\ \hline
4    & Adult Medical-Surgical & Emergency              & Circulatory physiology \\ \hline
5    & Emergency              & Adult Medical-Surgical & Neurology              \\ \hline
6    & Electrophysiology      & Electrophysiology      & Ambulatory Surgery     \\ \hline
7    & Circulatory physiology & Circulatory physiology & Adult Medical-Surgical \\ \hline
8    & Neurology              & Neurology              & Pediatric Cardiology   \\ \hline
9    & Unknown                & Unknown                & Emergency              \\ \hline
10   & Pediatric Cardiology   & Pediatric Cardiology   & Unknown                \\ \hline
\end{tabular}\npnoround
}
\end{table*}

In Table \ref{compare}, we show the top ranked services, i.e., which services are the best wayports, under three conditions. Condition 1 uses only those trails in the health record dataset that are unbroken. It can be viewed as the true rank. For condition 2, we changed the temporal resolution to 9 days, which resulted in 297 broken trails. We ignored the broken points and got the second inverted betweenness rank without recovery. Finally, condition 3, we applied our framework to these broken trails and got the third rank after order recovery. As shown in this table, the true rank is very similar with the rank after location order recovery. The Spearman rank-order correlation coefficient\cite{zwillinger1999crc} between these two ranks is 0.976 which is much larger than the correlation between the truth rank and the rank without recovery(0.491).

%% file: content/6_conclusion.tex
\section{Conclusion and Discussion}
In this paper, we studied location order recovery for trails with low temporal resolution. It is a missing data issue that prevents us knowing the correct moving order. After defining the concept of broken point, we examined the wide existence of broken points in a real health record dataset. 

In our experiments, we designed two strategies to create artificial broken points on two datasets. Hence we can know the ground truth for these broken points. Experiments on these two datasets have shown the effectiveness of our framework. We showed that adding BEGIN/END nodes in the original trails can effectively capture the beginning and ending information. Trail partitioning can dramatically increase the location order recovery accuracy by overcoming the re-entering issue. After this two procedures, we find the transition route with the highest probability. Through the distribution of the length of continuous broken points, we showed that an exact algorithm is feasible for a large portion of them. However, there are still a few broken points with extra large sizes which cannot be handled efficiently by an exact algorithm.  Hence we proposed utilizing an ant colony algorithm to recover the true location orders in those super long broken points. In our case study, we showed that the location recovery framework can effectively capture the important locations and the result is highly correlated with the rank got from trails under high temporal resolution.



As with any research there are limitations. One such limitation is that there are some error patterns frequently occurring in our experiment which cannot be dealt with efficiently using our framework. For example,our approach cannot distinguish location sequences $ABAACA$ and $ACAABA$ because the joint probability products are the same for these two sequences. A second limitation is that our framework is based on the first-order Markov transition probability between locations. Using higher order Markov transition probability may prove effective at improving the recovery accuracy\cite{shamshad2005first}. The third limitation is the Markov assumption.  Future work might consider how to model trails without the Markov assumption.

This paper presented a framework for sequence recover when there portions of a sequence are missing or collapsed due to lack of temporal resolution.  Such a framework has applicability to many scenarios involving object movements like human mobility research and traffic flow study. Despite limitations, this framework provides a powerful approach for accurately inferring order despite missing data.  Future work should explore the application of this method to diverse data sets. An advantage of the proposed framework is that researchers can plug a variety of algorithms designed for the asymmetric traveling salesman problem into it. This makes it both more generalizable and gives it the potential for further performance improvements. We anticipate that such improvements, in conjunction with this framework, will reduce the fragility of machine learning techniques that are predicated on knowing sequences and so pave the way for improved prediction.



%% file: content/7_acknowledge.tex
\section*{Acknowledgment}
The authors would like to thank all members of the Center for Computational Analysis of Social and Organizational Systems for their valuable advice. These findings were derived from data supplied by Columbia University Medical Center.  The author's thank Jacqueline Merrill, Professor of Nursing in Biomedical Informatics, and her research team for their advice and insight.

%% file: main.bbl
\begin{thebibliography}{10}
\providecommand{\url}[1]{#1}
\csname url@samestyle\endcsname
\providecommand{\newblock}{\relax}
\providecommand{\bibinfo}[2]{#2}
\providecommand{\BIBentrySTDinterwordspacing}{\spaceskip=0pt\relax}
\providecommand{\BIBentryALTinterwordstretchfactor}{4}
\providecommand{\BIBentryALTinterwordspacing}{\spaceskip=\fontdimen2\font plus
\BIBentryALTinterwordstretchfactor\fontdimen3\font minus
  \fontdimen4\font\relax}
\providecommand{\BIBforeignlanguage}[2]{{%
\expandafter\ifx\csname l@#1\endcsname\relax
\typeout{** WARNING: IEEEtran.bst: No hyphenation pattern has been}%
\typeout{** loaded for the language `#1'. Using the pattern for}%
\typeout{** the default language instead.}%
\else
\language=\csname l@#1\endcsname
\fi
#2}}
\providecommand{\BIBdecl}{\relax}
\BIBdecl

\bibitem{gonzalez2008understanding}
M.~C. Gonzalez, C.~A. Hidalgo, and A.-L. Barabasi, ``Understanding individual
  human mobility patterns,'' \emph{Nature}, vol. 453, no. 7196, pp. 779--782,
  2008.

\bibitem{wilcove2008going}
D.~S. Wilcove and M.~Wikelski, ``Going, going, gone: is animal migration
  disappearing,'' \emph{PLoS Biol}, vol.~6, no.~7, p. e188, 2008.

\bibitem{lv2015traffic}
Y.~Lv, Y.~Duan, W.~Kang, Z.~Li, and F.-Y. Wang, ``Traffic flow prediction with
  big data: a deep learning approach,'' \emph{IEEE Trans. Intell. Transp.
  Syst.}, vol.~16, no.~2, pp. 865--873, 2015.

\bibitem{tatem2006global}
A.~J. Tatem, D.~J. Rogers, and S.~Hay, ``Global transport networks and
  infectious disease spread,'' \emph{Advances in parasitology}, vol.~62, pp.
  293--343, 2006.

\bibitem{gruhl2004information}
D.~Gruhl, R.~Guha, D.~Liben-Nowell, and A.~Tomkins, ``Information diffusion
  through blogspace,'' in \emph{Proceedings of the 13th international
  conference on World Wide Web}.\hskip 1em plus 0.5em minus 0.4em\relax ACM,
  2004, pp. 491--501.

\bibitem{hagerstrand1968innovation}
T.~Hagerstrand \emph{et~al.}, ``Innovation diffusion as a spatial process.''
  \emph{Innovation diffusion as a spatial process.}, 1968.

\bibitem{chen2011discovering}
Z.~Chen, H.~T. Shen, and X.~Zhou, ``Discovering popular routes from
  trajectories,'' in \emph{Data Engineering (ICDE), 2011 IEEE 27th
  International Conference on}.\hskip 1em plus 0.5em minus 0.4em\relax IEEE,
  2011, pp. 900--911.

\bibitem{pei2001prefixspan}
J.~Pei, J.~Han, B.~Mortazavi-Asl, H.~Pinto, Q.~Chen, U.~Dayal, and M.-C. Hsu,
  ``Prefixspan: Mining sequential patterns efficiently by prefix-projected
  pattern growth,'' in \emph{icccn}.\hskip 1em plus 0.5em minus 0.4em\relax
  IEEE, 2001, p. 0215.

\bibitem{srikant1996mining}
R.~Srikant and R.~Agrawal, \emph{Mining sequential patterns: Generalizations
  and performance improvements}.\hskip 1em plus 0.5em minus 0.4em\relax
  Springer, 1996.

\bibitem{han2000freespan}
J.~Han, J.~Pei, B.~Mortazavi-Asl, Q.~Chen, U.~Dayal, and M.-C. Hsu, ``Freespan:
  frequent pattern-projected sequential pattern mining,'' in \emph{Proceedings
  of the sixth ACM SIGKDD international conference on Knowledge discovery and
  data mining}.\hskip 1em plus 0.5em minus 0.4em\relax ACM, 2000, pp. 355--359.

\bibitem{hung2009mining}
C.-C. Hung, C.-W. Chang, and W.-C. Peng, ``Mining trajectory profiles for
  discovering user communities,'' in \emph{Proceedings of the 2009
  International Workshop on Location Based Social Networks}.\hskip 1em plus
  0.5em minus 0.4em\relax ACM, 2009, pp. 1--8.

\bibitem{dijkstra1959note}
E.~W. Dijkstra, ``A note on two problems in connexion with graphs,''
  \emph{Numerische mathematik}, vol.~1, no.~1, pp. 269--271, 1959.

\bibitem{luo2013finding}
W.~Luo, H.~Tan, L.~Chen, and L.~M. Ni, ``Finding time period-based most
  frequent path in big trajectory data,'' in \emph{Proceedings of the 2013 ACM
  SIGMOD International Conference on Management of Data}.\hskip 1em plus 0.5em
  minus 0.4em\relax ACM, 2013, pp. 713--724.

\bibitem{stark1989high}
H.~Stark and P.~Oskoui, ``High-resolution image recovery from image-plane
  arrays, using convex projections,'' \emph{JOSA A}, vol.~6, no.~11, pp.
  1715--1726, 1989.

\bibitem{donoho1993nonlinear}
D.~L. Donoho, ``Nonlinear wavelet methods for recovery of signals, densities,
  and spectra from indirect and noisy data,'' in \emph{In Proceedings of
  Symposia in Applied Mathematics}.\hskip 1em plus 0.5em minus 0.4em\relax
  Citeseer, 1993.

\bibitem{pan2016image}
J.~Pan, X.~Yang, H.~Cai, and B.~Mu, ``Image noise smoothing using a modified
  kalman filter,'' \emph{Neurocomputing}, vol. 173, pp. 1625--1629, 2016.

\bibitem{Huang2017}
B.~Huang and K.~M. Carley, ``On predicting geolocation of tweets using
  convolutional neural networks,'' in \emph{Social, Cultural, and Behavioral
  Modeling: 10th International Conference, SBP-BRiMS 2017}, 2017, pp. 281--291.

\bibitem{qian2017probabilistic}
Y.~Qian, J.~Tang, Z.~Yang, B.~Huang, W.~Wei, and K.~M. Carley, ``A
  probabilistic framework for location inference from social media,''
  \emph{arXiv preprint arXiv:1702.07281}, 2017.

\bibitem{acar2011scalable}
E.~Acar, D.~M. Dunlavy, T.~G. Kolda, and M.~M{\o}rup, ``Scalable tensor
  factorizations for incomplete data,'' \emph{Chemometrics and Intelligent
  Laboratory Systems}, vol. 106, no.~1, pp. 41--56, 2011.

\bibitem{dai2009subspace}
W.~Dai and O.~Milenkovic, ``Subspace pursuit for compressive sensing signal
  reconstruction,'' \emph{IEEE Transactions on Information Theory}, vol.~55,
  no.~5, pp. 2230--2249, 2009.

\bibitem{dempster1977maximum}
A.~P. Dempster, N.~M. Laird, and D.~B. Rubin, ``Maximum likelihood from
  incomplete data via the em algorithm,'' \emph{Journal of the royal
  statistical society. Series B (methodological)}, pp. 1--38, 1977.

\bibitem{zhang2017rate}
Y.~Zhang, W.~Wei, B.~Huang, K.~M. Carley, and Y.~Zhang, ``Rate: Overcoming
  noise and sparsity of textual features in real-time location estimation,'' in
  \emph{Proceedings of the 2017 ACM on Conference on Information and Knowledge
  Management}.\hskip 1em plus 0.5em minus 0.4em\relax ACM, 2017, pp.
  2423--2426.

\bibitem{merrill2015transition}
J.~A. Merrill, B.~Sheehan, K.~Carley, and P.~Stetson, ``Transition networks in
  a cohort of patients with congestive heart failure: A novel application of
  informatics methods to inform care coordination,'' \emph{Applied clinical
  informatics}, vol.~6, no.~3, pp. 548--564, 2015.

\bibitem{fischetti1992additive}
M.~Fischetti and P.~Toth, ``An additive bounding procedure for the asymmetric
  travelling salesman problem,'' \emph{Mathematical Programming}, vol.~53,
  no.~1, pp. 173--197, 1992.

\bibitem{cirasella2001asymmetric}
J.~Cirasella, D.~S. Johnson, L.~A. McGeoch, and W.~Zhang, ``The asymmetric
  traveling salesman problem: Algorithms, instance generators, and tests,'' in
  \emph{Workshop on Algorithm Engineering and Experimentation}.\hskip 1em plus
  0.5em minus 0.4em\relax Springer, 2001, pp. 32--59.

\bibitem{dorigo1997ant}
M.~Dorigo and L.~M. Gambardella, ``Ant colony system: a cooperative learning
  approach to the traveling salesman problem,'' \emph{IEEE Trans. Evol.
  Comput.}, vol.~1, no.~1, pp. 53--66, 1997.

\bibitem{zwillinger1999crc}
D.~Zwillinger and S.~Kokoska, \emph{CRC standard probability and statistics
  tables and formulae}.\hskip 1em plus 0.5em minus 0.4em\relax Crc Press, 1999.

\bibitem{shamshad2005first}
A.~Shamshad, M.~Bawadi, W.~W. Hussin, T.~Majid, and S.~Sanusi, ``First and
  second order markov chain models for synthetic generation of wind speed time
  series,'' \emph{Energy}, vol.~30, no.~5, pp. 693--708, 2005.

\end{thebibliography}
